\documentclass[prl,reprint,superscriptaddress,preprintnumbers]{revtex4-1}
\pdfoutput=1
\usepackage{amsfonts, amssymb, amsmath}
\usepackage{mathrsfs}
\usepackage{hyperref}
\usepackage{tikz}
\usepackage[table]{}
\usepackage{empheq}
\usepackage{physics}

\usepackage{tensor}

\usepackage{color}
\definecolor{dark-gray}{gray}{0.20}
\definecolor{gray}{gray}{0.30}
\definecolor{light-gray}{gray}{0.80}
\definecolor{dark-red}{rgb}{0.7,0,0}
\definecolor{dark-green}{rgb}{0.1,0.4,0}
\definecolor{dark-blue}{rgb}{0.3,0.3,0.7}
\definecolor{light-blue}{rgb}{0.8,0.8,1}
\definecolor{blue}{rgb}{0,0,1}
\definecolor{red}{rgb}{1,0,0}
\definecolor{green}{rgb}{0,1,0}

\usepackage{hyperref}
\hypersetup{
	colorlinks=true,
	linkcolor=dark-blue,
	citecolor=dark-red,
	urlcolor=dark-blue,
	linktoc=page
}

\def\Tr{{\rm Tr}\,}

\newcommand{\be}{\begin{equation}}
\newcommand{\ee}{\end{equation}}
\newcommand{\bea}{\begin{eqnarray}}
\newcommand{\eea}{\end{eqnarray}}

\newcommand{\U}{\text{U}}

\usepackage{tikz}

\usepackage{xcolor}

\def\U{{\rm U}}

\begin{document}

\title{Higher-Derivative Corrections and AdS$_5$ Black Holes}

\date{\today}

\author{Nikolay Bobev}

\affiliation{Instituut voor Theoretische Fysica, KU Leuven, Celestijnenlaan 200D, B-3001 Leuven, Belgium}

\author{Vasil Dimitrov}

\affiliation{Instituut voor Theoretische Fysica, KU Leuven, Celestijnenlaan 200D, B-3001 Leuven, Belgium}

\author{Valentin Reys}

\affiliation{Instituut voor Theoretische Fysica, KU Leuven, Celestijnenlaan 200D, B-3001 Leuven, Belgium}

\author{Annelien Vekemans}

\affiliation{Instituut voor Theoretische Fysica, KU Leuven, Celestijnenlaan 200D, B-3001 Leuven, Belgium}

\begin{abstract}
\noindent Using recent results in four-derivative 5d $\mathcal{N}=2$ minimal gauged supergravity, we evaluate the regularized on-shell action of the Euclidean solution in this theory that admits a Lorentzian continuation to an AdS$_5$ black hole with one electric charge and two angular momenta. We focus on the supersymmetric limit of this solution and employ holography to show that the result can be expressed purely in terms of angular momentum fugacities and the 't Hooft anomalies for the $\U(1)_{\rm R}$ R-symmetry of the dual 4d $\mathcal{N}=1$ SCFT. This holographic calculation is in perfect agreement with recent studies of the 4d $\mathcal{N}=1$ superconformal index ``on the second sheet''. We illustrate the utility of these results in two classes of 4d $\mathcal{N}=1$ holographic SCFTs that have 't Hooft anomalies with suitable large $N$ behavior that leads to non-trivial corrections at first subleading order in the $1/N$ expansion. We also explicitly calculate the Wald entropy of the black hole solution and delineate the leading four-derivative corrections to the Bekenstein-Hawking entropy.
\end{abstract}


\maketitle
\noindent\textbf{Introduction}--Exploring the AdS/CFT correspondence beyond the classical two-derivative supergravity approximation is a challenging endeavor that is bound to lead to important insights into quantum gravity. Supersymmetry is an invaluable tool that allows for quantitative access to physical observables on both sides of the duality. There have been recent important advances in this context using supersymmetric localization for 3d~$\mathcal{N}=2$ SCFTs and higher-derivative corrections in 4d supergravity. In particular, supersymmetric localization allows for the systematic calculation of the path integral of the SCFT on compact Euclidean manifolds order by order in the $1/N$ expansion, see~\cite{Pestun:2016zxk} for a review. For SCFTs arising from M2-branes, the leading and subleading terms at large $N$ can be reproduced using holography by studying the four-derivative corrections to 4d gauged supergravity~\cite{Zaffaroni:2019dhb,Benini:2015eyy,Bobev:2020egg,Bobev:2021oku}. Our goal in this work is to extend this success to 5d gauged supergravity and the dual 4d $\mathcal{N}=1$ SCFTs.

The QFT observable of interest here is the superconformal index which can be defined for any 4d $\mathcal{N}=1$ SCFT as the path integral of the theory on $S^1\times S^3$ with appropriate supersymmetric periodicity conditions \cite{Romelsberger:2005eg,Kinney:2005ej,Rastelli:2016tbz}. The large $N$ limit of the superconformal index for holographic SCFTs is subtle and it was recently understood that one needs to consider the analyticity properties of the index on the complex fugacity plane in order to find non-trivial large $N$ scaling indicative of a weakly coupled holographic bulk dual, see \cite{Hosseini:2017mds,Cabo-Bizet:2018ehj,Benini:2018ywd,Choi:2018hmj} and references thereof.  While the superconformal index contains detailed information about the spectrum of BPS states in the 4d $\mathcal{N}=1$ SCFT, it simplifies drastically in the large $N$ limit when one takes the fugacities to be on the ``second sheet'' \cite{Cassani:2021fyv}. It was argued in \cite{Cassani:2021fyv,GonzalezLezcano:2020yeb}, that the superconformal index on the second sheet, $\mathcal{I}$, takes the following form
\begin{align}\label{eq:CK2ndsheet}
&\log \mathcal{I} = \frac{(\omega_1+\omega_2+2\pi {\rm i} n_0)^3}{48\omega_1\omega_2} {\rm Tr} R^3 \\ 
& -\frac{(\omega_1+\omega_2+2\pi {\rm i}n_0)(\omega_1^2+\omega_2^2-4\pi^2)}{48\omega_1\omega_2} {\rm Tr} R +\log |G|\,,\notag
\end{align}
where $\omega_{1,2}$ are the two angular momentum fugacities, ${\rm Tr} R^3$ and ${\rm Tr} R$ are the cubic and linear 't Hooft anomalies for $\U(1)_{\rm R}$, the supersymmetry constraint that fixes the R-symmetry fugacity has already been implemented with $n_0=\pm 1$, and $|G|$ is the order of the Abelian one-form symmetry in the theory (if any). This result is valid up to terms of order $e^{-\ell_{3}/\beta}$ in the small $\beta$ limit, where $\beta$ is the circumference of $S^1$ and $\ell_3$ is the radius of the $S^3$, and includes the contribution from the supersymmetric Casimir energy \cite{Assel:2015nca,Bobev:2015kza}. Our main result in this paper is a supergravity calculation that reproduces the first two terms on the RHS of~\eqref{eq:CK2ndsheet}. This constitutes a non-trivial precision test of holography and provides strong independent evidence for the validity of the results in \cite{Cassani:2021fyv,GonzalezLezcano:2020yeb}.

For holographic SCFTs with a weakly coupled supergravity dual, ${\rm Tr} R^3$ is the leading term in the large $N$ expansion and is proportional to the dimensionless ratio $L^3/G_5$ between the scale $L$ of AdS$_5$ and the 5d Newton constant. Indeed, it was shown in \cite{Cabo-Bizet:2018ehj} how the first term on the RHS of \eqref{eq:CK2ndsheet} can be obtained holographically by calculating the regularized on-shell action of a class of supersymmetric Euclidean solutions of 5d minimal $\mathcal{N}=2$ gauged supergravity with an $S^1\times S^3$ boundary derived in \cite{Chong:2005da,Chong:2005hr}. In SCFTs with an appropriate large $N$ scaling the ${\rm Tr} R$ term in \eqref{eq:CK2ndsheet} should be captured by including four-derivative corrections to the supergravity theory. Recently the full 5d minimal $\mathcal{N}=2$ gauged supergravity action including the two possible four-derivative terms was derived in \cite{Bobev:2021qxx}, see also \cite{Liu:2022sew}. We show that evaluating the regularized four-derivative on-shell action for the CCLP solution of \cite{Chong:2005da,Chong:2005hr} in this four-derivative theory yields precise agreement with the ${\rm Tr} R^3$ and ${\rm Tr} R$ terms in \eqref{eq:CK2ndsheet}. We then employ this result and the first law of thermodynamics to calculate the Wald entropy of the Lorentzian AdS$_5$ black hole obtained by analytic continuation of the CCLP solution and comment on its relation to the superconformal index \footnote{After the first version of this manuscript appeared on the arXiv we became aware of \cite{Cassani:2022lrk} which has significant overlap with our work. Our results for the four-derivative on-shell action are in full agreement. However, we have now adopted the approach advocated in \cite{Cassani:2022lrk} for the calculation of the corrections to the Bekenstein-Hawking entropy. This leads to a modification of \eqref{eq:SWaldBPS} as compared to the first version of this manuscript.}.
\smallskip

\noindent\textbf{The Euclidean CCLP solution}--It was shown in~\cite{Chong:2005da,Chong:2005hr} that the following background, which we will refer to as the CCLP solution, solves the equations of motion of the two-derivative 5d minimal $\mathcal{N}=2$ gauged supergravity
\begin{equation}\label{eq:CCLPmetA}
    \begin{split}
        ds^2 =& \frac{\Delta_\eta \left[ \left( 1+\frac{r^2}{L^2} \right)\rho^2\,d\tau + 2{\rm i} q\nu \right]d\tau}{\Xi_a \Xi_b \rho^2} +\frac{2q\nu\omega}{\rho^2} \\&+\frac{f}{\rho^4}\left(\frac{{\rm i}\Delta_\eta\, d\tau}{\Xi_a\Xi_b} + \omega \right)^2 +\frac{\rho^2dr^2}{\Delta_r} + \frac{\rho^2d\eta^2}{\Delta_\eta} \\
        &+\frac{r^2+a^2}{\Xi_a}\sin^2\eta\,d\xi_1^2 + \frac{r^2+b^2}{\Xi_b}\cos^2\eta\,d\xi_2^2\,,\\
        A =& -\frac{\sqrt{3}q}{\rho^2}\left( \frac{ {\rm i}\Delta_\eta  d\tau}{\Xi_a \Xi_b} + \omega \right) -{\rm i} \alpha\,d\tau,
    \end{split}
\end{equation}
where we have defined
\begin{align}\label{eq:CCLPdef}
        &\nu = b\sin^2\eta\,d\xi_1 + a\cos^2\eta\,d\xi_2\,, \notag\\
        &\Delta_r = \frac{(r^2+a^2)(r^2+b^2)\left( 1+\frac{r^2}{L^2} \right) + q^2 + 2abq-2mr^2}{r^2}\,,\notag\\
        & \omega = \frac{a}{\Xi_a}\sin^2\eta\,d\xi_1 + \frac{b}{\Xi_b}\cos^2\eta\,d\xi_2\,,\\
        &\Delta_\eta = 1-\frac{a^2}{L^2}\cos^2\eta - \frac{b^2}{L^2}\sin^2\eta\,,\notag\\
        &\rho^2 = r^2 + a^2\cos^2\eta + b^2\sin^2\eta\,,\notag\\
        &f=2m\rho^2-q^2+\frac{2abq\rho^2}{L^2}\,, ~~ \Xi_a = 1-\frac{a^2}{L^2}\,, ~~ \Xi_b = 1-\frac{b^2}{L^2}\,.\notag
\end{align}
We present the solution in Euclidean signature and do not assume any reality properties of the four parameters $(m,q,a,b)$. The solution in Lorentzian signature can be obtained from the expressions above by the replacement $\tau \to {\rm i} t$ and describes a black hole in AdS$_5$ with one electric charge and two independent angular momenta. 

We are interested in studying the effects of the four-derivative corrections to 5d $\mathcal{N}=2$ gauged supergravity on the on-shell action and thermodynamic properties of this solution. These four-derivative corrections were studied recently in \cite{Bobev:2021qxx} where it was shown that they are controlled by two constant coefficients $c_{1,2}$, see \eqref{eq:L-corr} in the Appendix for the full bosonic Lagrangian of the supergravity theory. As discussed in \cite{Reall:2019sah,Melo:2020amq}, see also \cite{Gubser:1998nz,Caldarelli:1999ar} for earlier observations to this effect, if one is interested in corrections linear in $c_{1,2}$ it is not necessary to correct the two-derivative solution but it is sufficient to consider only the effects due to the corrected action. It is expected on very general grounds that $c_{1,2}$ are parametrically small corrections to the two-derivative action~\cite{Camanho:2014apa}. We will therefore use the results in \cite{Reall:2019sah,Melo:2020amq} as our starting point and work with the two-derivative CCLP solution~\footnote{Note that the logic here is different from the one used in \cite{Bobev:2020egg,Bobev:2021oku} where it was shown that every solution of the two-derivative minimal 4d $\mathcal{N}=2$ gauged supergravity also solves the equations of motion of the four-derivative theory without assuming that the coefficients $c_{1,2}$ are small.}.

The Euclidean solution caps off at the location of the outer black hole horizon obtained by solving $ \Delta_r(r_+) = 0$. The length $\beta$ of the Euclidean time circle is determined by requiring regularity of the metric in the limit $r \to r_{+}$ and reads
\begin{equation}\label{eq:betadef}
\beta  = \frac{2\pi r_+[(r_+^2 + a^2)(r_+^2 + b^2)+abq]}{r_+^4[1 + g^2(2r_+^2 + a^2 + b^2)] - (ab + q)^2} \,,
\end{equation}
where $g=1/L$ and the coordinate ranges of the Euclidean solution are 
\begin{equation}\label{eq:coordrange}
 \tau \in [0, \beta)\,, ~  r \in [r_+ , \infty)\,, ~  \eta \in [0, \tfrac{\pi}{2})\,, ~  \xi_{1,2} \in [0, 2\pi)\,.
\end{equation}
The Euclidean CCLP solution is supersymmetric if one constrains the parameters as
\begin{align}\label{eq-susy-limit}
    q & = \frac{m}{1 + ag + bg} \,.
\end{align}
It is important to stress that this supersymmetric Euclidean solution does not in general admit a Lorentzian continuation to a supersymmetric and causal black hole solution, see \cite{Cabo-Bizet:2018ehj} for a more detailed discussion. To obtain the BPS Lorentzian black hole found in \cite{Gutowski:2004ez,Chong:2005da,Chong:2005hr} one has to impose the additional relation 
\begin{equation}\label{eq-mt}
\frac{mg}{(a + b)(1 + ag)(1 + bg)(1 + ag + bg)} = 1 \,,
\end{equation}
together with the inequality $a + b + abg > 0$.

To properly understand the four-derivative corrections to the thermodynamic properties of this gravitational solution one needs to carefully compute the conserved charge, angular momentum, and entropy. We will discuss some aspects of these calculations below. For the chemical potentials dual to these thermodynamic variables one can simply employ the fact that we are working with an uncorrected two-derivative solution and use the temperature $T=1/\beta$, with $\beta$ in \eqref{eq:betadef}, as well as the following
angular and electric chemical potentials \cite{Cabo-Bizet:2018ehj}
\begin{align}\label{eq-potentials}
    \begin{aligned}
        \Omega_1 & = \frac{a(r_+^2 + b^2)(1 + g^2r_+^2) + bq}{(r_+^2 + a^2)(r_+^2 + b^2) + abq} \,, \\ 
        \Omega_2 &= \frac{b(r_+^2 + a^2)(1 + g^2r_+^2) + aq}{(r_+^2 + a^2)(r_+^2 + b^2) + abq} \,, \\
        \Phi & = \frac{\sqrt{3} q r_+^2}{(r_+^2 + a^2)(r_+^2 + b^2) + abq}  \,.
    \end{aligned}
\end{align} 
It is important to stress that imposing regularity of the gauge field at $r=r_+$, i.e. demanding that $A_\mu A^\mu$ is finite, fixes the pure gauge constant $\alpha$ in \eqref{eq:CCLPmetA} as $\alpha = -\Phi$.

Studying thermodynamics for supersymmetric black holes is subtle due to their finite entropy and vanishing temperature. Here we will follow the approach of~\cite{Cabo-Bizet:2018ehj}, see also \cite{Bobev:2020pjk}, and use the supersymmetric Euclidean solution as a crutch to calculate finite gravitational observables in the supersymmetric limit. This means that we will study the two-parameter family of supersymmetric Euclidean solutions obtained by imposing the constraint~\eqref{eq-susy-limit}. The supersymmetric temperature and chemical potentials are then obtained from~\eqref{eq:betadef} and~\eqref{eq-potentials} by imposing the constraint~\eqref{eq-susy-limit} and will be denoted by a superscript $s$, while the relevant quantities for the supersymmetric Lorentzian black hole solution will be denoted by a $*$ superscript and are obtained by additionally imposing~\eqref{eq-mt}. To make a connection with the fugacities in the $S^1\times S^3$ supersymmetric partition function of the dual SCFT it was shown in~\cite{Cabo-Bizet:2018ehj} that it is useful to define the following angular and electric gravitational fugacities
\begin{align}\label{eq-new-susy-potentials}
    \begin{aligned}
        \omega_1^s & \equiv \beta^s(\Omega_1^s - \Omega_1^*) = \frac{\pi(ag-1)(b - {\rm i}r_+)}{\Xi} \,,\\
        \omega_2^s & \equiv \beta^s(\Omega_2^s - \Omega_2^*) = \frac{\pi(bg-1)(a - {\rm i}r_+)}{\Xi} \,, \\
        \varphi^s & \equiv \beta^s(\Phi^s - \Phi^*) = \frac{ \sqrt{3}\, \pi(a - {\rm i}r_+)(b - {\rm i}r_+)}{\Xi}\,,
    \end{aligned}
\end{align}
where $\Xi\equiv r_+(1 + ag + bg) + \frac{{\rm i}g}{2}(r_*^2 - 3r_+^2)$ and $r_*$ is the value of $r_{+}$ when both constraints in~\eqref{eq-susy-limit} and~\eqref{eq-mt} are imposed. These new chemical potentials satisfy the linear constraint
\begin{align}\label{eq-susy-constraint2}
    \omega_1^s + \omega_2^s -  \frac{\sqrt{3}}{L}\, \varphi^s & = 2\pi {\rm i} \,,
\end{align}
which is the supergravity analog of the linear relation between the SCFT fugacities that is already implemented in \eqref{eq:CK2ndsheet}. The fact that the supergravity fugacities add up to an imaginary number reflects the analytic continuation to the second sheet used in \cite{Cassani:2021fyv}.
\smallskip

\noindent\textbf{HD corrections and the on-shell action}--According to the standard holographic dictionary the supergravity dual of the logarithm of the superconformal index in~\eqref{eq:CK2ndsheet} should be equal to the regularized on-shell action of the CCLP solution.  To compute this action we use the four-derivative supergravity Lagrangian in \eqref{eq:L-corr} and regulate the UV divergences by employing background subtraction. More specifically, we calculate the difference between the on-shell action of the CCLP solution for general values of the parameters (using the coordinate ranges in~\eqref{eq:coordrange}) and the on-shell action of empty AdS$_5$ in global coordinates. We note that the same holographic regularization scheme was used in the two-derivative on-shell action analysis in~\cite{Chen:2005zj,Cabo-Bizet:2018ehj}. The on-shell action calculation is arduous and was performed with the help of {\tt Mathematica}~\footnote{Our code for the calculations that lead to the results presented in this paper is based on the {\tt xAct} package and can be found at \url{https://github.com/waskou/SolutionsX}.}. The final result for the regularized four-derivative on-shell action however is compact and simple and reads
\begin{equation}\label{eq:Ireg}
\begin{split}
    I_{\text{reg}} &= \frac{(\varphi^s)^3}{\omega_1^s\omega_2^s}\left[\frac{\pi }{12\sqrt{3}G_5}-\frac{2\pi^2(c_1+6c_2)}{L^2}\right] \\
    & \qquad\qquad +\frac{\varphi^s[(\omega_1^s)^2+(\omega_2^s)^2-4\pi^2]}{\omega_1^s\omega_2^s}2\pi^2 c_1.
 \end{split}
\end{equation}
To compare this calculation with the field theory result in~\eqref{eq:CK2ndsheet} we need to invoke the holographic map between field theory and gravitational quantities. The relation between the supergravity parameters and the 't Hooft anomaly coefficients in the dual SCFT was worked out in \cite{Bobev:2021qxx} and reads~\footnote{The parameter $L$ we use in this work is related to the $\ell$ used in (2.54) of \cite{Bobev:2021qxx} by $L = \ell + 16\pi\sqrt{3} G_5 c_2/\ell$.}
\begin{align}\label{eq:TrR2TrR}
        &\Tr R^3 =\frac{16(5\mathfrak{a}-3\mathfrak{c})}{9} = \frac{4\pi L^3}{9G_5} - \frac{32\sqrt{3}\pi^2}{3}L(c_1+6c_2)\,, \notag\\
        &\Tr R = 16(\mathfrak{a}-\mathfrak{c})= -32\sqrt{3}\pi^2L c_1\,.
\end{align}
As discussed in \cite{Cabo-Bizet:2018ehj}, the supersymmetric angular fugacities $\omega_{1,2}^s$ in \eqref{eq-new-susy-potentials} should be identified with $\omega_{1,2}$ in~\eqref{eq:CK2ndsheet}. Using all this, together with the relation between the supergravity fugacities in \eqref{eq-susy-constraint2},  we find a perfect agreement between the four-derivative regularized on-shell action in \eqref{eq:Ireg} and the first two terms on the RHS of~\eqref{eq:CK2ndsheet} for $n_0=-1$ \footnote{The fact that our supergravity result yields $n_0=-1$ has to do with the choice of conventions for the sign of the electric charge of the CCLP solution.}. This agreement constitutes a precision test of holography beyond the two-derivative supergravity approximation.

The field theory result in~\eqref{eq:CK2ndsheet} applies to all 4d $\mathcal{N}=1$ SCFTs. Similarly, the four-derivative 5d supergravity on-shell action in~\eqref{eq:Ireg} is  universal and should apply to all holographic SCFTs with a weakly coupled bulk dual. It is important however to keep in mind that not all holographic SCFTs have an appropriate large $N$ scaling of $\Tr R$ which ensures that the four-derivative supergravity couplings $c_{1,2}$ do not vanish. For instance, as discussed in \cite{Bobev:2021qxx}, in $\mathcal{N}=4$ SYM one has $\Tr R=0$, while in the well-known $Y^{(p,q)}$ quiver gauge theories $\Tr R \sim N^{0}$ and thus we find that $c_{1,2}$ vanish in these models. As a result, one probably needs to study loop effects in supergravity to capture the $\Tr R=0$ term in~\eqref{eq:CK2ndsheet}. Two notable examples where the 5d Lagrangian in \eqref{eq:L-corr} captures the effects of ${\rm Tr} R$ are the 4d $\mathcal{N}=2$ F-theory models discussed in \cite{Aharony:1998xz,Aharony:2007dj} and the 4d $\mathcal{N}=1$ class $\mathcal{S}$ SCFTs that arise from M5-branes wrapped on a Riemann surface \cite{Maldacena:2000mw,Gaiotto:2009gz,Bah:2011vv,Bah:2012dg}. In both of these examples ${\rm Tr} R$ scales as $N$ and, as shown in  \cite{Bobev:2021qxx}, $c_{1,2}$ do not vanish.

To be concrete, the $\U(1)_{\rm R}$ 't Hooft anomalies for the 4d $\mathcal{N}=2$ SCFTs arising from D3-branes at F-theory singularities are \cite{Aharony:1998xz,Aharony:2007dj}
\begin{equation}\label{eq:Fthanom}
\begin{split}
\Tr R^3 &= \frac{8\Delta}{9} N^2+\frac{4(\Delta-1)}{9} N + \frac{2}{27}\,,\\
 \Tr R &= 4(1-\Delta) N + \frac{2}{3}\,,
\end{split}
\end{equation}
where $N$ is the number of D3-branes and the rational number $\Delta$ specifies the global flavor symmetry of the SCFT and is related to the number of D7-branes, $n_7$, used in the construction. It is given by $\Delta = \frac{12}{12-n_7}$ with $n_7$ taking values in the set $\{2,3,4,6,8,9,10\}$. 

For the theories arising from wrapped M5-branes on a smooth compact Riemann surface $\Sigma_{\mathfrak{g}}$ of genus $\mathfrak{g}>1$ the 't Hooft anomalies were computed in \cite{Bah:2011vv,Bah:2012dg} and the result to leading and subleading order at large $N$ reads
\begin{align}\label{eq:Mthanom}
\Tr R^3 &= \frac{2(\mathfrak{g}-1)}{27z^2}\left[9z^2-1+(3z^2+1)^{\frac{3}{2}}\right]N^3\notag\\
& -\frac{\mathfrak{g}-1}{9z^2}\left[(\sqrt{3z^2+1}-1)(2+3z^2)-3z^2\right]N+\ldots\,,\notag\\
 \Tr R &= \frac{\mathfrak{g}-1}{3}\left[4-\sqrt{3z^2+1}\right]N+\ldots\,.
\end{align}
Here $N$ is the number of M5-branes and the rational parameter $z$ specifies the choice of partial topological twist on $\Sigma_{\mathfrak{g}}$ for the theory on the worldvolume of the M5-branes. We can use the anomalies \eqref{eq:Fthanom} and \eqref{eq:Mthanom}, together with the supergravity relation in \eqref{eq:TrR2TrR}, to find the large $N$ behavior of the superconformal index in these two classes of models. We emphasize that for these models the large $N$ behavior of the superconformal index has not been studied by ``standard methods'' and the only available large $N$ results are the application of the universal QFT formula in~\eqref{eq:CK2ndsheet} and the supergravity results presented above.
\smallskip

\noindent\textbf{Wald entropy}--In a higher-derivative diffeomorphism covariant theory of gravity the entropy of a black hole solution can be computed using Wald's formalism \cite{Wald:1993nt}. Given the Lagrangian density of the $d$-dimensional theory, $\mathcal{L}$, the Wald entropy can be found through the following integral
\begin{equation}\label{eq:SWdef}
S_{\rm W} = -2\pi \int_{\Sigma} d^{d-2}x \sqrt{\gamma}\,\frac{\delta \mathcal{L}}{\delta R_{\mu\nu\rho\sigma}}\,\epsilon_{\mu\nu}\epsilon_{\rho\sigma} \, ,
\end{equation}
where $\Sigma$ is a bifurcate Killing horizon with induced metric $\gamma_{\mu\nu}$, $\epsilon_{\mu\nu}$ is the binormal to $\Sigma$ defined as $\kappa\,\epsilon_{\mu\nu} = \nabla_\mu V_\nu$ with $V^\mu$ the null Killing vector generating the horizon and $\kappa$ the surface gravity. As emphasized in \cite{Tachikawa:2006sz}, the Wald procedure has to be modified in the presence of gravitational Chern-Simons terms. Applying the results of \cite{Tachikawa:2006sz} to the $A\wedge R \wedge R$ term in the four-derivative Lagrangian~\eqref{eq:L-corr} we find the following additional contribution to the Wald entropy
\begin{equation}\label{eq:SCSW}
S^{\text{CS}}_{\rm W} = -2\pi\int_\Sigma d^3x\,\varepsilon^{\mu\nu\rho}\,\epsilon^\sigma{}_\lambda\,\Gamma^\lambda{}_{\sigma\mu}\,F_{\nu\rho} \, .
\end{equation}
Using \eqref{eq:SWdef} and \eqref{eq:SCSW} and the Lagrangian in \eqref{eq:L-corr} we find that the Wald entropy of any black hole solution in the four-derivative 5d $\mathcal{N}=2$ minimal supergravity can be found by computing the integral in \eqref{eq:SWaldApp}. The result of this calculation for the two-derivative CCLP solution is discussed in the Appendix. Unfortunately this result is not compatible with the first law of thermodynamics which strongly suggests that one should first correct the CCLP solution using the four-derivative equations of motion and then evaluate the Wald entropy.

In the absence of the explicit corrections to the CCLP background, one can use the four-derivative gravitational on-shell action and the first law of thermodynamics as a shortcut to obtain the entropy of the corrected black hole solution as follows. Choosing an ensemble where the fugacities $\{\beta,\Phi,\Omega_{1,2}\}$ are kept fixed, varying the on-shell action with respect to the quantities in \eqref{eq:betadef} and \eqref{eq-potentials} yields the corrected angular momenta $J_{1,2} = -\frac{1}{\beta}\frac{\partial I}{\partial \Omega_{1,2}}$, electric charge $Q = -\frac{2L}{\sqrt{3}\beta}\frac{\partial I}{\partial \Phi}$ and energy $E = \frac{\partial I}{\partial \beta} + \Omega_1 J_1 + \Omega_2 J_2 + \frac{\sqrt{3}}{2L}\Phi Q$ \footnote{This shortcut to calculating the four-derivative charges and entropy was proposed in \cite{Cassani:2022lrk} after the first version of the present paper appeared on the arXiv.}. The normalization of $Q$ is chosen for later convenience. In the BPS limit, these conserved quantities obey the linear relation $2LE = 2(J_1 + J_2) + 3Q$. Using this and the first law of thermodynamics, the entropy of the four-derivative BPS black hole reads:
\begin{align}\label{eq:SWaldBPS}
S_{\text{W}}^{{\rm BPS}} &= \pi \sqrt{3 Q^2 - 8\mathfrak{a}(J_1+J_2)-\frac{16\mathfrak{a}(\mathfrak{a}-\mathfrak{c})(J_1-J_2)^2}{Q^2 - 2\mathfrak{a}(J_1+J_2)}}\,.
\end{align}
We write this formula as a non-linear expression in terms of the black hole charges and the conformal anomalies of the dual SCFT. This is inspired by the two-derivative result in \cite{Kim:2006he} but is somewhat misleading. All quantities in \eqref{eq:SWaldBPS} should be linearized in the HD parameters $c_{1,2}$ to obtain the four-derivative black hole entropy in terms of the parameters of the CCLP solution. The explicit expressions for the charges and black hole entropy to linear order in $c_{1,2}$ are unwieldy and can be found in the online repository \cite{Note3}. We have also checked that the linearized form of~\eqref{eq:SWaldBPS} follows from a Legendre transform of the on-shell action in \eqref{eq:Ireg} with respect to $\{\varphi^s,\omega_{1,2}^s\}$ subject to the constraint in \eqref{eq-susy-constraint2}, and then taking the BPS limit.

\smallskip

\noindent\textbf{Discussion}--In this work we established a precise agreement between the leading and first subleading correction in the large $N$ limit of the superconformal index of a holographic 4d $\mathcal{N}=1$ SCFT and the on-shell action of the dual CCLP supergravity solution evaluated in the four-derivative 5d $\mathcal{N}=2$ minimal gauged supergravity. Importantly, for the subleading correction to be non-vanishing in supergravity we need an appropriate large $N$ scaling of the linear and cubic 't Hooft anomalies. In particular, our results imply that in \cite{Melo:2020amq} no correction to the black hole entropy was observed simply because ${\rm Tr} R=0$ in $\mathcal{N}=4$ SYM and thus $c_{1,2}$ vanish.

Our results also point to a number of important open questions. It is desirable to put on a more solid footing the thermodynamic properties of the CCLP solution in the four-derivative theory of interest here. As discussed above, the chemical potentials remain the same as their two-derivative values and we have calculated the four-derivative corrections to the on-shell action. Calculating the corrections to the entropy, charge, mass, and angular momentum is however non-trivial since it involves finding the corrections to the CCLP solution due to the four-derivative terms in the action as well as a proper treatment of holographic renormalization in the presence of higher-derivative terms. It will certainly be most interesting to compute these quantities explicitly and in particular check the validity of the quantum statistical relation and the first law of black hole thermodynamics~\cite{Gibbons:1976ue,Gibbons:2004ai}, which we have assumed to be valid when computing the black hole entropy in \eqref{eq:SWaldBPS}. This is important for properly deriving and understanding the black hole entropy microscopically in string or M-theory. In this context it is also desirable to understand how to use supergravity or string theory to account for the higher order terms in the $1/N$ expansion of the superconformal index as well as for the exponentially suppressed corrections. We hope that our results here are a useful step in that direction.

%%%%%%%%%%%%%%
%\smallskip

\noindent\textbf{Acknowledgements}--We are grateful to D.~Cassani, A.~Charles, J.~Hong, D.~Martelli, A.~Zaffaroni, and especially K.~Hristov for useful discussions. We also thank the authors of \cite{Cassani:2022lrk} for the useful communication after the first version of this work appeared on the arXiv. We are supported in part by an Odysseus grant G0F9516N from the FWO and by the KU Leuven C1 grant ZKD1118 C16/16/005. VD and AV are also supported by doctoral fellowships from the Research Foundation - Flanders (FWO).
%%%%%%%%%%%%%%%%%
%
\bibliography{HD-M5-BH}
%
%%%%%%%%%%%%%%%
%
\newpage
\noindent\textbf{Appendix}--As shown in \cite{Bobev:2021qxx} the four-derivative bosonic Lagrangian of 5d minimal $\mathcal{N}=2$ gauged supergravity takes the following form 
\begin{widetext}
\begin{align}
\label{eq:L-corr}
e^{-1}\mathcal{L}_{4\partial} =&\; -\Bigl[\frac{1}{\kappa^2} + \frac{(5\,c_1 + 24\,c_2)g^2}{2\sqrt{3}}\Bigr]R +\Bigl[\frac{1}{4\kappa^2} + \frac{7(5\,c_1 - 12\,c_2)g^2}{24\sqrt{3}}\Bigr]F_{ab}^2 \nonumber - \Bigl[\frac{12g^2}{\kappa^2} + \frac{1}{\sqrt{3}}\,(25\,c_1 + 156\,c_2)\,g^4\Bigr] \nonumber \\[1mm] 
&- \frac{{\rm i}}{12\sqrt{3}}\,\Bigl[\frac{1}{\kappa^2} - \frac{3\sqrt{3}(c_1 + 6\,c_2)}{2}g^2\Bigr]e^{-1}\varepsilon^{\mu\nu\rho\sigma\tau}A_\mu F_{\nu\rho}F_{\sigma\tau} - \frac{{\rm i}c_1}{16}e^{-1}\varepsilon^{\mu\nu\rho\sigma\tau}A_\mu R_{\nu\rho}{}^{\lambda\epsilon}R_{\sigma\tau\lambda\epsilon}  \\[1mm]
&- \frac{(2\,c_1 - 3\,c_2)}{24\sqrt{3}}R\,F_{ab}^2 + \frac{5c_1}{4\sqrt{3}}R^{ab}F_{ac}F_b{}^c - \frac{\sqrt{3}c_1}{16}R_{abcd}F^{ab}F^{cd} - \frac{(c_1 + 6\,c_2)}{8\sqrt{3}}\,R^2 + \frac{c_1}{2\sqrt{3}}R_{ab}^2 - \frac{\sqrt{3}c_1}{8}(R_{abcd})^2 \nonumber\\[1mm]
&- \frac{5\sqrt{3}}{64}\,c_1\,F^{ab}F_{a}{}^cF_b{}^d F_{cd} + \frac{(61\,c_1 - 6\,c_2)}{1152\sqrt{3}}F_{ab}^2\,F_{cd}^2 + \frac{\sqrt{3}c_1}{2}(\nabla_a F_{bc})(\nabla^{[a}F^{b]c}) + \frac{\sqrt{3}c_1}{2}F_{ab}\nabla^b\nabla_c F^{ac} \nonumber \\[1mm] 
&- \frac{{\rm i}c_1}{8}e^{-1}\varepsilon^{\mu\nu\rho\sigma\tau} F_\mu{}^\lambda F_{\sigma\tau}\Bigl(\frac32\,\nabla_\nu F_{\lambda\rho} - \nabla_\lambda F_{\nu\rho}\Bigr) - \frac{3{\rm i}c_1}{32}e^{-1}\varepsilon^{\mu\nu\rho\sigma\tau} F_{\mu\nu}F_{\rho\sigma}\nabla^\lambda F_{\lambda\tau} + \mathcal{O}(c_i^2) \, . \nonumber
\end{align}
\end{widetext}
We have written the Lagrangian in Euclidean signature, used $e$ to denote the square root of the determinant of the metric,  and have defined the $\varepsilon$ symbol as $\varepsilon^{01234} = \varepsilon_{01234} = 1$. In addition to the metric, the only bosonic field in the theory is the $\U(1)$ gauge field $A_{\mu}$ with field strength $F_{\mu\nu}$ and we use conventions where $\kappa^2 = 16 \pi G_5$ and the supergravity gauge coupling is related to the AdS$_5$ scale $L$ used in \eqref{eq:CCLPmetA}, \eqref{eq:CCLPdef} as $g=1/L$. We emphasize that \eqref{eq:L-corr} was derived by working to linear order in the small parameters $c_{1,2}$ determining the four-derivative invariants. When the couplings $c_{1,2}$ vanish, the Lagrangian above reduces to the one for the two-derivative 5d minimal $\mathcal{N}=2$ gauged supergravity. 

Around \eqref{eq:SWdef} we explained how to implement the Wald entropy calculation for our four-derivative supergravity theory. The result of this lengthy analysis for the Lagrangian in \eqref{eq:L-corr} is given by the following expression
\begin{widetext}
\begin{align}\label{eq:SWaldApp}
S_{\rm W} = - 2\pi\int_\Sigma d^3x\sqrt{\gamma}\Bigl[&\Bigl(\frac{1}{\kappa^2} + \frac{5c_1 + 24c_2}{2\sqrt{3}}g^2 + \frac{2c_1 - 3c_2}{24\sqrt{3}}F_{ab}^2 + \frac{c_1 + 6c_2}{4\sqrt{3}}R\Bigr)\eta^{[a[c}\eta^{d]b]}-\frac{c_1}{4\sqrt{3}}\Bigl(5F^{eg}F^{f}{}_g + 4 R^{ef}\Bigr)\eta^{[a[c}\eta^{d]}{}_e \eta^{b]}{}_f  \notag\\
&\; + \frac{\sqrt{3}\,c_1}{16} \Bigl(F^{ab}F^{cd} + 4R^{abcd}\Bigr)\Bigl]\epsilon_{ab}\,\epsilon_{cd} -\frac{c_1 \pi}{2}\int_\Sigma d^3x  \,\varepsilon^{\mu\nu\rho}\,\epsilon^\sigma{}_\lambda\,\Gamma^\lambda{}_{\sigma\mu} F_{\nu\rho} \, .
\end{align}
\end{widetext}
The last term on the RHS of \eqref{eq:SWaldApp} is the contribution from the $A\wedge R \wedge R$ term in the supergravity action. It is straightforward but quite tedious to evaluate the Wald entropy for the general non-supersymmetric two-derivative CCLP black hole solution in \eqref{eq:CCLPmetA}, \eqref{eq:CCLPdef}. We performed the calculation using {\tt Mathematica} and found explicitly the Wald entropy for the uncorrected black hole solution. Unfortunately this result for the Wald entropy does not agree with the entropy computed using the first law of thermodynamics and presented in \eqref{eq:SWaldBPS}. This implies that in order to find the correct Wald entropy one needs to find the correction to the CCLP solution due to the presence of the four-derivative terms in the supergravity action.

\end{document}